\documentclass[12pt]{article}
\begin{document}

\title{{\bf Huge Quantum Gravity Effects \\in the Solar System}}

\author{
Don N. Page
\thanks{Internet address:
don@phys.ualberta.ca}
\\
Theoretical Physics Institute\\
Department of Physics, University of Alberta\\
Room 238 CEB, 11322 -- 89 Avenue\\
Edmonton, Alberta, Canada T6G 2G7
}

\date{2010 May 17}

\maketitle
\large
\begin{abstract}
\baselineskip 25 pt

Normally one thinks of the motion of the planets around the Sun as a highly
classical phenomenon, so that one can neglect quantum gravity in the Solar
System.  However, classical chaos in the planetary motion amplifies quantum
uncertainties so that they become very large, giving huge quantum gravity
effects.  For example, evidence suggests that Uranus may eventually be
ejected from the Solar System, but quantum uncertainties would make the
direction at which it leaves almost entirely uncertain, and the time of its
exit uncertain by about a billion billion years.  For a time a billion
billion years from now, there are huge quantum uncertainties whether Uranus
will be within the Solar System, within the Galaxy, or even within causal
contact of the Galaxy.

\end{abstract}

\normalsize

\baselineskip 23 pt

\newpage 

The motion of planets in the Solar System is often regarded as the paradigm
of a completely deterministic system, say in the nonrelativistic model of
Newtonian gravitation for a spherical Sun and spherical planets.  Of course
it is known that a deterministic classical model is only an approximation
for an underlying quantum description, but one might estimate that quantum
uncertainties should be quite negligible for the Solar System.

For example, if one has an object with mass $M$ and no external forces on
it, one may put its center of mass into a minimum-uncertainty (gaussian)
quantum wavefunction with rms position uncertainty $\Delta x$ and rms
momentum uncertainty $\Delta p = \hbar/\Delta x$.  Then after a time $t$,
the motion due to the rms momentum uncertainty will combine with the
original rms position uncertainty to give a total rms position uncertainty
$\sqrt{\Delta x^2 + [(\hbar t)/(M\Delta x)]^2}$.  For fixed $t$, this has
the minimal value $\sqrt{2\hbar t/M}$.  If we take $t$ to be the age of the
Solar System, 4.6 billion years, and $M$ to be the mass of the Earth,
$5.972\times 10^{24}$ kilograms, then the minimal rms position uncertainty
is $1.2\times 10^{-27}$ meters, which for practical purposes is quite
negligible.  That is, it appears that if one put a planet into a
minimal-uncertainty wavepacket with suitable $\Delta x$, the spreading would
be negligible over the lifetime of the Solar System.

One can also consider the quantum effect of sunlight scattering off the
Earth, which would be expected to give the Earth an rms momentum roughly
equal to the momentum of each photon multiplied by the square root of the
number of scattered photons.  Over the age of the Solar System, this would
give a position uncertainty of the order of ten nanometers ($10^{-8}$ m),
much larger than the uncertainties without scattered sunlight, but again in
practice quite negligible.

Thus it appears that the quantum fluctuations from the Uncertainty Principle
applied to the center of mass of the Earth and from the quantum fluctuations
in the scattering of sunlight from the Earth are always quite tiny.  Such
considerations seem to suggest that quantum effects are very small for the
motion of planets in the Solar System.

Nevertheless, there are other effects that can make quantum effects become
enormous.  In particular, classical chaos in the Solar System
\cite{SusWis,Laskar,SusWis2,Peters,DunQui,Las2,MurHol,
MalHolIto,LecFraHolMur} can exponentially amplify quantum uncertainties
\cite{BerZas,BerBal,ZurPaz,Zur} to become huge uncertainties in the
positions of the planets.  Because these effects involve both quantum
uncertainties and the uncertain gravitational fields produced by the Sun and
planets at their uncertain positions, they are truly large quantum gravity
effects.  Here I shall argue that one set of such uncertainties is the time
and velocity at which Uranus is likely to be ejected from the Solar System.

As summarized in \cite{LecFraHolMur}, it was recognized that a point-mass
Newtonian model with three or more masses could exhibit both chaotic and
regular motion, but up to the 1980s, it was thought that the actual
conditions for the Sun and planets of our Solar System would result in
regular motion.  Therefore, it was a surprise in 1988 when Sussmann and
Wisdom \cite{SusWis} showed that Pluto's motion was chaotic, with a Lyapunov
time (the timescale for exponential growth of uncertainties) of about 20
Myr.  The next year Laskar \cite{Laskar} found that the entire Solar System
was chaotic, with a Lyapunov time of about 5 Myr, as confirmed in more
detail in 1992 by Sussman and Wisdom \cite{SusWis2}, who also found that the
outer Solar System by itself was chaotic with a Lyapunov time of about 7
Myr.  Nevertheless, in 1994 Laskar \cite{Las2} integrated the motion of all
the planets except Pluto for 25 Gyr and found no ejections or collisions, so
it seems that nothing drastic may happen to the present planets until the
Sun swells up into a red giant and engulfs the inner Solar System out
through the earth 7.6 Gyr from now \cite{SchSmi}.  If the Sun did not do
that, Laskar found evidence supporting the idea that Mercury might be
removed by a collision with the Sun or Venus on a timescale of roughly 2000
Gyr, far later than the Sun would actually swallow it.

Despite the fact that the finite future lifetime of the Sun does not seem to
give enough time for the inner Solar System to undergo a drastic transition
(e.g., by losing a planet) before it gets destroyed by the Sun's red giant
phase, there can still be huge uncertainties in the relative positions of
the planets.  Zurek \cite{Zur}, expanding upon ideas of
\cite{BerZas,BerBal,ZurPaz}, has noted that an upper bound on the timescale
for which the evolution of a classically chaotic system becomes flagrantly
non-classical (huge amplified quantum uncertainties) is the Lyapunov time
multiplied by the logarithm of the action of the system in units of
$\hbar$.  If we take the action of the Solar System as the angular momentum
of the planetary system, which is \cite{Allen4} $3.148\times 10^{43}$ kg
m$^2$ s$^{-1} = 2.985\times 10^{77}\hbar$, its logarithm is 178.393, so if
one uses a Lyapunov time of 5 Myr, one gets that quantum uncertainties
should be amplified to huge values by 900 Myr, which is about 20\% of the
present age of the Solar System.  Therefore, one would expect that even if
one measured the position and momentum as precisely at possible, the
relative positions of the planets would have uncertainties of the order of
an astronomical unit after another billion years or so.  (The distances
between the planets and the Sun are not likely to be uncertain by nearly so
much within this timescale of gigayears, but the angles between the planets
and the Sun would be expected to have quantum uncertainties of the order of
radians.)

On a longer timescale, the ultimate fate of the outer Solar System seems
likely to have even much greater uncertainties. It has recently been shown
\cite{MurHol,MalHolIto,LecFraHolMur} that overlapping three-body resonances
among Jupiter, Saturn, and Uranus give a Lyapunov time of 5--10 Myr and an
escape time for Uranus of the order of $10^{18}$ years, say from Uranus'
orbit becoming perturbed to pass close to Saturn and gain enough kinetic
energy by the slingshot effect to escape the gravitational pull of the Sun. 
The ratio of the escape time to the Lyapunov time, $10^{11}$ or so, is of
the order of a billion times the logarithm of the action, so the details of
the escape will have huge quantum uncertainties.

In particular, the time at which Uranus escapes from the Solar System is
likely to have quantum uncertainties of the same order of magnitude as the
escape time itself, of the order of a billion billion ($10^{18}$) years. 
With an escape velocity of the order of the orbital velocity of Uranus,
$2\times 10^{-5} c$, and with a Galactic size of the order of 25 kiloparsecs
or close to $10^5$ light years, it would take less than $10^{10}$ years for
Uranus to move very far across the Galaxy from where the Sun is.  Therefore,
in $10^{18}$ years, Uranus would have time to make hundreds of millions of
orbits around the Galaxy relative to the Sun.  Furthermore, the relaxation
time for an object to change its orbit in the Galaxy significantly has been
given \cite{Allen3} in the solar neighborhood as $2.6\times 10^6$ years
multiplied by the cube of the velocity in kilometers per second, which gives
nearly $10^9$ years for the orbital velocity of Uranus.  Thus in a time
comparable to the escape time from the Solar System of $10^{18}$ years,
Uranus would have time for something like a billion ($10^9$) significant
changes in its orbit in the Galaxy.  This should be plenty of time for
Uranus to gain a significant probability of escaping not only from the Solar
System, but also from the entire Galaxy.

If the present value of the dark energy persists, so that universe continues
to expand exponentially at a timescale of the order of $10^{10}$ years, by a
proper time equal to the expected escape time from the Solar System of
$10^{18}$ years, Uranus will not only have a significant probability of
escaping from both the Solar System and the Galaxy, but also of crossing the
cosmological event horizon, so that no signals can thereafter get back from
Uranus to the rest of the Solar System.  Therefore, whether or not Uranus
can even in principle be seen from another planet like Saturn at the
expectation value of its escape time will have enormous quantum gravity
uncertainties.

Thus quantum gravity effects in the motions of the planets give a huge
uncertainty in the time at which Uranus will escape from the Solar System,
as well as in the escape direction.  At a time corresponding to the
expectation value of this escape time, the position of Uranus could, with
significant probabilities, be either within the Solar System, almost
anywhere within the Galaxy, or even beyond the cosmological event horizon. 
Although quantum gravity effects within the Solar System may be small within
a human lifetime, over the timescale needed for a planet like Uranus to
escape, quantum gravity effects are enormous.

\end{document}